\documentclass[preprint,superscriptaddress,aip,onecolumn,12pt]{revtex4-1}
\usepackage{graphicx}
\usepackage{dcolumn}
\usepackage{bm}
\usepackage{amsmath}
\usepackage{hyperref}
\newcommand{\lgchptr}[1]{\vspace{0.5cm}{\large{\textbf{#1}}}\\}
\newcommand{\sublgchptr}[1]{\vspace{0.25cm}{\normalsize\textbf{#1}}\\}
\begin{document}

\title{Fast Purcell-enhanced single photon source in 1,550-nm telecom band from a resonant quantum dot-cavity coupling.}
\author{Muhammad Danang Birowosuto}
\affiliation{NTT Basic Research Laboratories, NTT Corporation, 3-1 Morinosato Wakamiya, Atsugi, Kanagawa 243-0198, Japan}
\author{Hisashi Sumikura}
\affiliation{NTT Basic Research Laboratories, NTT Corporation, 3-1 Morinosato Wakamiya, Atsugi, Kanagawa 243-0198, Japan}
\author{Shinji Matsuo}
\affiliation{NTT Photonics Laboratories, NTT Corporation, 3-1 Morinosato Wakamiya, Atsugi, Kanagawa 243-0198, Japan}
\author{Hideaki Taniyama}
\affiliation{NTT Basic Research Laboratories, NTT Corporation, 3-1 Morinosato Wakamiya, Atsugi, Kanagawa 243-0198, Japan}
\author{Peter J. van Veldhoven}
\affiliation{COBRA Research Institute, Eindhoven University of Technology, Postbus 513, 5600 MB Eindhoven, The Netherlands}
\author{Richard N\"otzel}
\affiliation{COBRA Research Institute, Eindhoven University of Technology, Postbus 513, 5600 MB Eindhoven, The Netherlands}
\affiliation{Present Address: Institute for Systems based on Optoelectronics and Microtechnology (ISOM), ETSI Telecommunication,
Technical University of Madrid, Ciudad Universitaria s/n, 28040 Madrid, Spain}
\author{Masaya Notomi$^{*}$}
\altaffiliation{Corresponding author.}
\affiliation{NTT Basic Research Laboratories, NTT Corporation, 3-1 Morinosato Wakamiya, Atsugi, Kanagawa 243-0198, Japan}
\date{\today}% It is always \today, today,

\begin{abstract}
\textbf{High-bit-rate nanocavity-based single photon sources in the 1,550-nm
telecom band are challenges facing the development of fibre-based
long-haul quantum communication networks. Here we report a very fast
single photon source in the 1,550-nm telecom band, which is achieved by
a large Purcell enhancement that results from the coupling of a single InAs 
quantum dot and an InP photonic crystal nanocavity. At a resonance, the spontaneous
emission rate was enhanced by a factor of 5 resulting a record fast
emission lifetime of 0.2 ns at 1,550 nm. We also demonstrate that this emission
exhibits an enhanced anti-bunching dip. This is the first realization of
nanocavity-enhanced single photon emitters in the 1,550-nm telecom band.
This coupled quantum dot cavity system in the telecom band thus provides
a bright high-bit-rate non-classical single photon source that offers appealing
novel opportunities for the development of a long-haul quantum telecommunication
system via optical fibres.}
\end{abstract}

\maketitle

In terms of realising a long-distance quantum telecommunication network, a single photon source in the telecom band around 1,550 nm is crucial since this band offers the lowest transmission loss in the fibre-based quantum communication network. Increasing the bit rate of single photon emission is also important for maintaining a high transferred bit rate after a long transmission. Therefore, there is a strong demand for high-repetition-rate single photon sources in the telecom band for quantum key distribution (QKD) applications \cite{Gisin2002,Knill2001} and also for quantum repeater networks. Single self-assembled semiconductor quantum dots (QDs) are attractive single photon sources since they are fast, bright, photostable, and scalable on-demand sources of single photons \cite{Shields2007,Lounis2005}. Some interesting characteristics were also reported with these QDs such as improvements in the photon extraction efficiency \cite{Strauf2007,Claudon2010}, the suppression of multiphoton events \cite{Ho2009}, the photon indistinguishability \cite{Santori2002a}, coherent control by strongly driven resonance fluorescence \cite{Flagg2009}, the temporal modulation of triggered single photon \cite{Rakher2011}, and an electrically driven single photon \cite{Yuan2002,Intallura2007}. Despite these research activities, studies of QDs at the telecom wavelengths are still very limited \cite{Zinoni2006, Intallura2007, Takemoto2007, Miyazawa2008a, Takemoto2010a, Dorenbos2010}, and only a few experiments have demonstrated QKD applications in the 1,550-nm telecom band.

To realise high-repetition-rate single photon sources, we should enhance the radiative recombination rate, which can be achieved by coupling with a high-\emph{Q} nanocavity via the Purcell effect \cite{Vahala2003}. This effect is very promising because high-quality nanocavities have become available owing to recent progress made on photonic crystals \cite{Notomi2010}. This is especially important in the 1,550-nm telecom band since the transition dipole moment of QDs in this band is generally smaller than that of QDs at shorter wavelengths \cite{Gong2010e}. There was a notable report on photoluminescence (PL) measurements obtained from the coupling between a QD and a cavity \cite{Dalacu2010} but there has been no report on cavity-enhanced single photon emission in the 1,550-nm telecom band. For single photon emission without employing cavity enhancement, the highest repetition rate and the shortest radiative recombination time in this band are 20 MHz and 1.12 ns, respectively \cite{Takemoto2010a}. Chauvin \emph{et al.} observed a moderate enhancement of a factor of 1.5 in telecom-band QDs coupled with extended slow light modes in 2D photonic crystals, but the lifetime was still 1.2 ns and they did not confirm single photon emission \cite{Chauvin2009}.

In this letter, we demonstrate a single photon source at 1,557 nm with a radiative recombination time of 0.2 ns (over five times faster than the previous record \cite{Takemoto2010a}), which shows clear antibunching characteristics. Our approach is to embed InAs QDs within width-modulated line-defect cavities \cite{Kuramochi2006,Tanabe2007} in InP photonic crystals, which realises strong Purcell enhancement of the radiation rate. In addition, we employ the biexciton transition of a QD  \cite{Thompson2001}, which has a much higher radiative transition rate than that of a conventional exciton transition. As a result of these approaches, we have achieved a very fast single photon emission in the telecom band. This single photon source is suitable for a QKD experiment with a high modulation rate. This system is also promising for a high-bit-rate entangled photon pair source, which is needed for quantum repeaters.

\begin{figure}[h]
\includegraphics[width=6in]{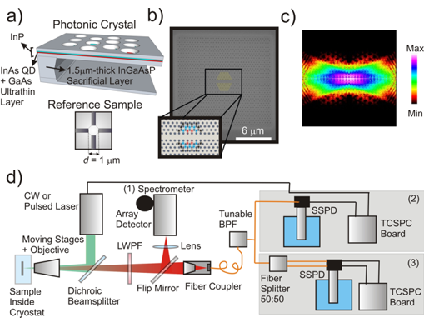}
\caption{\label{figure1}Details of samples and schematic of experimental set-up. \textbf{(a)} Structures of air-bridge photonic crystals and an octagonal flat-topped mesa with a diagonal length \emph{d}. InAs QDs and single ultrathin GaAs layer were grown between two 110-nm-thick InP layers. \textbf{(b)} Scanning electron micrograph of the sample and the design of the nanocavity in the line defect of an InP photonic crystal (inset). The cavity is shown as a yellow area and the holes shown with red, blue and black arrows are shifted by 9, 6 and 3 nm, respectively. \textbf{(c)} Vertical magnetic field $H_{y}$ profile of the resonant mode of the cavity for the $x-z$ plane in a finite-difference time-domain (FDTD) calculation. \textbf{(d)} Schematic of the experimental set-up for (1) photoluminescence (PL), (2) time-resolved PL and (3) second-order correlation measurements. The first and third experiments were performed with continuous wave (CW) excitation at 532 nm while the second experiment was recorded with pulsed laser excitation at 800 nm. LWPF, long-wave-pass filter; BPF, band-pass filter; SSPD, superconducting single photon detector; TCSPC, time-correlated single photon counting.}
\end{figure}

\lgchptr{Results}
For our experiment, we employed InAs/InP QDs grown by metalorganic vapor phase epitaxy in which an ultrathin GaAs interlayer is inserted beneath an InAs layer. This interlayer was introduced to ensure the fidelity of the intense emission of the QDs in the 1,550-nm region \cite{Anantathanasarn2005}. Two kinds of QD samples, a nanocavity in a photonic crystal and a reference sample were fabricated from an InAs/InP QD wafer by electron-beam lithography and high-resolution dry etching, see Fig. \ref{figure1}(a). The fabrication process is similar to that used in our previous reports for InP-based photonic crystals \cite{Nozaki2010,Matsuo2010}, see Methods for details. For the nanocavity, we adapt a width-modulated line-defect cavity in InP photonic crystals, which were reported in Ref. \cite{Kuramochi2006,Tanabe2007}, see Fig. \ref{figure1}(b) for the design and the scanning electron micrograph of the sample, respectively. The photonic crystal has an area of 14.3 $\times$ 15.4 $\mu$m$^{2}$ with a lattice constant \emph{a} of 440 nm. Fig. \ref{figure1}(c) shows the simulated magnetic field from the fundamental cavity mode of the line-defect cavity using the three-dimensional finite difference time domain (FDTD) method assuming a slab index \emph{n} of 3.17, a slab thickness \emph{t} of 224 nm, and a hole radius \emph{r} of 100 nm. The quality factor ($Q_{cal}$) and the modal volume ($V$) calculated for this design are $\sim$8,000 and 0.16 $\mu$m$^{3}$, respectively. The reference sample does not contain a cavity and consists of an octagonal flat-topped mesa with a span or a diagonal length \emph{d} of 1 $\mu$m.

A schematic of our experimental set-up is shown in Fig. \ref{figure1}(d). The QD in a sample inside a cryostat was optically excited using continuous wave (CW) excitation at 532 nm or 5-ps pulses from a 80 MHz mode-locked titanium sapphire laser at 800 nm. A free-space excitation technique was applied through a 0.42-numerical-aperture near-infrared microscope objective, by which the beam diameter was estimated to be $d \approx$ 3 $\mu$m. The fraction of the spontaneously emitted photons that coupled to radiation modes was collected through the same microscope objective and filtered with a dichroic beam splitter and a long-wave-pass filter (LWPF). To measure the QD PL spectrum, we directed the emission into a grating spectrometer with a cooled InGaAs array and a resolution of $\approx$ 0.1 nm. For time-resolved PL and second-order correlation measurements, the emission was focussed to a single mode fibre, which acted as a confocal pinhole for the spatial selection. A single QD signal was spectrally selected using a tunable bandpass filter (BPF) with a full-width at half maximum (FWHM) of 1 nm. For time resolved PL measurements, we directly coupled the filtered emission from the tunable BPF to one channel of a superconducting single photon detector (SSPD). Finally, we connected the SSPD and the excitation pulse signals to one input and a synchronization input channel of a time-correlated single photon counting (TCSPC) board, respectively. For the second-order correlation measurements, the emission photons after the tunable BPF were divided by a 50:50 fibre splitter and then directed into two channels of the SSPD, where detection events were used as the start and stop signals for the TCSPC.

\begin{figure}[h]
\includegraphics[width=6in]{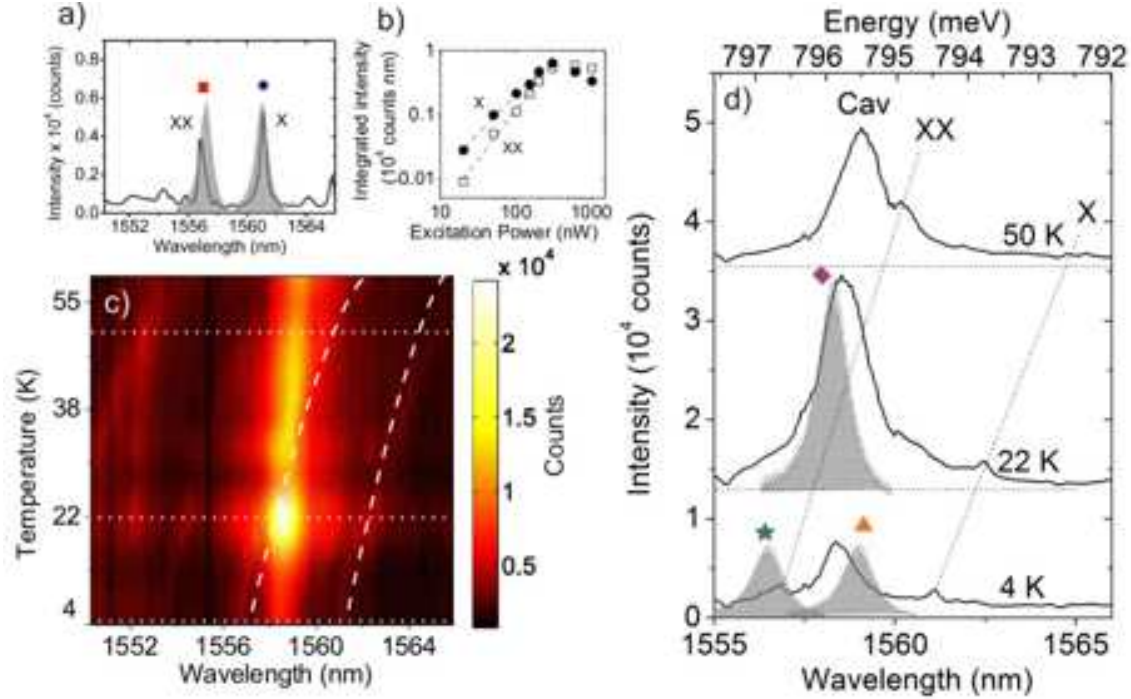}
\caption{\label{figure2}Photoluminescence spectra of InAs quantum dots in a reference sample and a cavity. \textbf{(a)} PL spectrum of a quantum dot in a reference sample taken with a grating spectrometer and an InGaAs array for an integration time of 20 s. The exciton and biexciton are denoted by \emph{X} and \emph{XX}, respectively. \textbf{(b)} Integrated intensities of the \emph{X} and \emph{XX} peaks as a function of CW laser power. Dotted lines are linear fits to the data. Both (a) and (b) were measured at 4 K. Dashed lines are linear fits to the data. \textbf{(c)} Density plot of the PL spectra of a quantum dot in a cavity obtained by scanning temperatures between 4 and 60 K and integrating 20 s at each point as a function of temperature (T). The dashed lines are a guide to the shift of the \emph{X} and \emph{XX} emission lines with temperature. \textbf{(d)} The single PL spectrum corresponds to the dotted lines in \textbf{(c)} with three different temperatures. The grey-shaded area with colored symbols shows the filter ranges for the time-resolved luminescence and second-order correlation measurements. All the spectra were measured with 200 nW of 532-nm-CW-excitation power.}
\end{figure}

Fig. \ref{figure2}(a) shows the PL spectrum of an InAs QD obtained from a reference sample in the 1,550-1,570 nm wavelength range \cite{Anantathanasarn2005}. The spectrum exhibits an excitonic (\emph{X}) line and a biexcitonic (\emph{XX}) line that peak at 1,556.8 and 1,561.1 nm, respectively. These peaks are similar to the \emph{X} and \emph{XX} lines reported by Cade \emph{et al.} \cite{Cade2006}. As also reported previously \cite{Cade2006}, the integrated intensities exhibit almost ideal linear and quadratic behavior for the \emph{X} and \emph{XX} lines, respectively, see Fig. \ref{figure2}(b) (details in Fig. S1, Supplementary Information). We observe the same \emph{X} and \emph{XX} emission lines for a single QD in a line-defect cavity (Fig. S2, Supplementary Information). The PL spectra for the QD cavity sample as a function of temperature (\emph{T}) are shown as a contour plot in Fig. \ref{figure2}(c) and the details of the spectra at \emph{T} = 4, 22 and 50 K are shown in Fig. \ref{figure2}(d). To record the PL spectra, we used an excitation power of 200 nW, which is still below the intensity saturation.  The \emph{X} and \emph{XX} emission lines at \emph{T} = 50 K are shifted by about 4 nm towards a longer wavelength than those at \emph{T} = 4 K (Fig. S3, Supplementary Information). For the cavity emission, we determine the experimental \emph{Q} ($Q_{exp}$) $\sim$ 2,000. The difference between the $Q_{exp}$ and the $Q_{cal}$ values may arise from fabrication errors such as air-hole diameter and position fluctuations. Temperature scanning revealed a \emph{XX}-cavity emission crossover at \emph{T} = 22 K. We assign the QD emission at \emph{T} = 4 K as being out of resonance with the cavity emission whereas that at \emph{T} = 22 K is on resonance. The PL intensity of the cavity at \emph{T} = 22 K is enhanced compared with that at \emph{T} = 4 K. After subtracting the integrated intensity of the cavity at \emph{T} = 4 K, the integrated intensity of the cavity at \emph{T} = 22 K is 6 times larger than that of the \emph{XX} line at \emph{T} = 4 K.

\begin{figure}[h]
\includegraphics[width=3.5in]{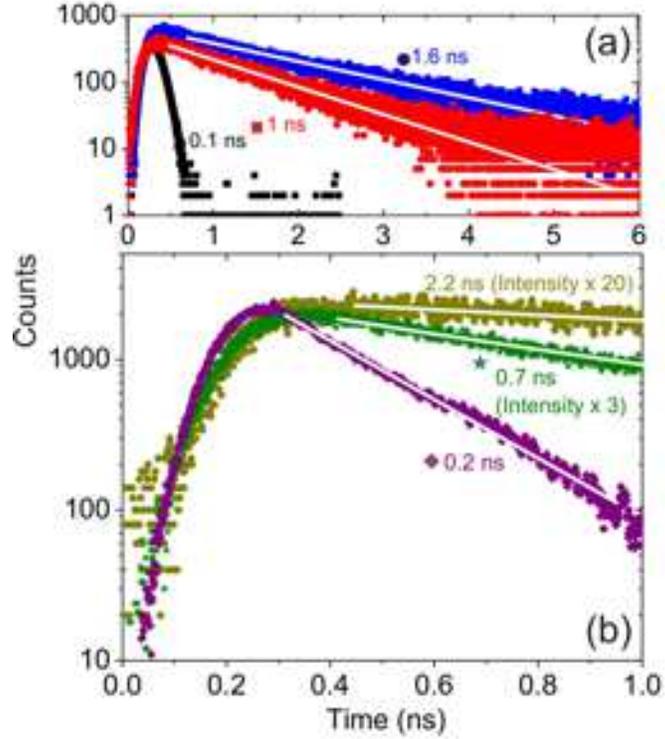}
\caption{\label{figure3}Time-resolved photoluminescence of a single QD. \textbf{(a)} Time-resolved PL of \emph{X} (blue circle) and \emph{XX} (red square) in a reference sample at \emph{T} = 4 K and \textbf{(b)} \emph{XX} in a cavity at \emph{T} = 4 K (green star) and zero detuning, \emph{T} = 22 K (purple diamond). The single exponential fit is shown by the solid white lines. The black squares and golden circles represent the instrumental response function of 0.1 ns and the inhibited emission of \emph{XX} in a photonic band gap of 2.2 ns at \emph{T} = 4 K, respectively. Other colored symbols correspond to the filter ranges in Fig. \textbf{\ref{figure2}}. All curves were subtracted from the background level.}
\end{figure}

Time-resolved PL was recorded on the \emph{X} and \emph{XX} lines for the reference sample and is shown in Fig. \ref{figure3}(a). The exponential fits to the time-resolved emissions of the \emph{X} and \emph{XX} lines in the reference sample show time constants of $\tau_{X,ref}$ = 1.6 ns and $\tau_{XX,ref}$ = 1.0 ns, respectively. Note that the latter is apparently faster than the \emph{X} emission. The 1.6-ns of $\tau_{X,ref}$ is a typical radiative lifetime of the \emph{X} recombination in InAs/InP QD systems \cite{Takemoto2007,Miyazawa2008a,Dorenbos2010}. We expect that $\tau_{XX,ref} \sim 0.5 \tau_{X,ref}$ since the radiative recombination of any of the two \emph{X} forming \emph{XX} will result its disappearance \cite{Thompson2001,Narvaez2005}. Additionally, our extracted $\tau_{X,ref}$ and $\tau_{XX,ref}$ are similar to those predicted for InAs/InP QD using atomic pseudopotential calculations \cite{Gong2010e}.

Fig. \ref{figure3}(b) shows the time-resolved emissions of \emph{XX} for the QD cavity sample at \emph{T} = 4 K (out of resonance) and \emph{T} = 22 K (on resonance). The total instrumental response function of 0.1 ns is also presented. It is apparent that when the \emph{XX} transition is on resonance with the cavity at \emph{T} = 22 K, the emission lifetime $\tau_{XX,cav,res}$ is greatly shortened to 0.2 ns. This result shows that $\frac{1/\tau_{XX,cav,res}}{1/\tau_{XX,ref}}$ = 5. We also show the decay curve of the XX emission in another photonic crystal sample where the XX emission lies completely within the photonic band gap at \emph{T} = 4 K, see Fig. \ref{figure3}(b) with details in Fig. S4, Supplementary Information. Note that the emission lifetime is elongated to 2.2 ns. This observation of the inhibition of the $\emph{XX}$ emission in the photonic band gap unambiguously proves that the observed fast decay curve is not due to a fabrication-induced non-radiative decay contribution. Therefore, the fivefold reduction in the lifetime must be caused by the cavity enhancement of the spontaneous emission rate. Compared with the previous reports \cite{Takemoto2007,Miyazawa2008a,Takemoto2010a,Chauvin2009}, this result is the fastest photon emission from QDs in the telecom band. We can also calculate the Purcell factor, $F_{p} = \frac{3}{4{\pi{}}^2}\left(\frac{\lambda}{n}\right)^{3}\frac{Q_{exp}}{V}$, where $\lambda$ is the cavity wavelength. With $Q_{exp}$ = 2,000 and $V$ = 0.16 $\mu$m$^{3}$, we found $F_{p}$ = 112.5. This value is about 22.5 times larger than $\frac{1/\tau_{XX,cav,res}}{1/\tau_{XX,ref}}$, but this is possible considering the spectral or/and spatial misalignment between the single QD and the cavity, and the polarization orientation of the cavity mode. Additionally, the discrepancy between the experimental emission rate enhancement and the $F_{p}$ is similar to that previously reported for InAs QDs in GaAs photonic crystals, in which the spatial and the spectral misalignments were estimated \cite{Badolato2005,Englund2005,Chang2006a}.

\begin{figure}[h]
\includegraphics[width=3.5in]{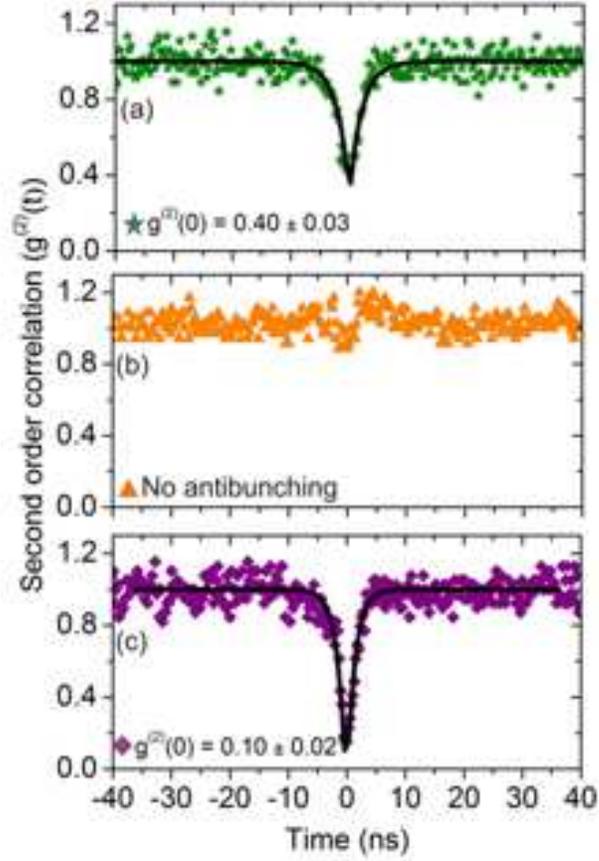}
\caption{\label{figure4}Single photon emission by a single QD in a cavity. Normalized histogram of a second-order correlation measurement $g^{(2)}(\tau)$ after background substraction for \textbf{(a)} \emph{XX} (green star), \textbf{(b)} cavity (orange triangle) at \emph{T} = 4 K, and \textbf{(c)} \emph{XX} at zero detuning, \emph{T} = 22 K (purple diamond). The colored symbols correspond to the filter range in Fig. \textbf{\ref{figure2}}.}
\end{figure}

The system was then arranged to measure the second-order correlation function, $g^{(2)}(\tau)$, for light emitted from \emph{XX}, see Fig. \ref{figure4}. Both channels of the SSPD send electrical pulses into a TCSPC module, which builds a histogram of the second-order correlation function. The signal from one detector was delayed by 60 ns to make it possible to see the negative time delay information. All the histograms were recorded with a 200 nW CW excitation power for integration times of 6,800 s and a time bin of 0.256 ns. The solid curve in Fig. \ref{figure4} is a fit to the standard expression $g^{(2)}(\tau) = 1-\left[[1-g^{(2)}(0)]\exp(-\mid{\tau}\mid/\tau_{tot})\right]$, where $\tau_{tot}$ is a time constant assigned to a combination of the emission lifetime and the inverse pumping rate, and $g^{(2)}(0) = 0$ is for a perfect antibunching (that is, the single photon state). To take account of the limited setup resolution, we convoluted the fit with a detector time-response function of a Gaussian distribution with an FWHM of 0.2 ns. Fig. \ref{figure4}(a) shows $g^{(2)}(\tau)$ for the \emph{XX} emission with the cavity at \emph{T} = 4 K (out of resonance), where the \emph{XX} emission exhibits partial antibunching with $g^{(2)}(0) = 0.40 \pm 0.03$ and $\tau_{tot} = 1.6 \pm 0.1$ ns. Fig. \ref{figure4}(b) shows $g^{(2)}(\tau)$ for the cavity emission at the same temperature, where we observed no antibunching behavior thus showing that the cavity emission is caused by classical background emissions. At zero detuning, as shown in Fig. \ref{figure4}(c), at \emph{T} = 22 K, the second-order correlation function exhibits clear antibunching behavior with $g^{(2)}(0) = 0.10 \pm 0.02$ and $\tau_{tot} = 1.2 \pm 0.1$ ns. This $g^{(2)}(0)$ is not very different from the $g^{(2)}(0)$ of $0.12 \pm 0.02$ obtained without the background subtraction (Fig. S5, Supplementary Information). We attribute the improvement in the antibunching compared with that in Fig. \ref{figure4}(a) to the reduction in the relative contribution from the background emissions caused by strong Purcell effect. This result clearly shows that the very fast \emph{XX} emission we observed in Fig. \ref{figure3}(b) is unambiguously non-classical light, namely it is in the single photon state. This directly shows that we achieved a single photon emitter with a 0.2 ns lifetime in the telecom wavelength band. In addition, our result shows that the time width $\tau_{tot}$ is $1.2 \pm 0.1$ ns, which is longer than the $\tau_{XX,cav,res}$ observed in Fig. \ref{figure3}(b). However, this is by far the shortest time width observed in a $g^{(2)}(0)$ measurement in the 1,550-nm telecom band. The previous $g^{(2)}(0)$ measurements were reported using gated InGaAs avalanche photodiodes (APDs) and the width was limited by the 50-ns time bin \cite{Takemoto2010a}.

\lgchptr{Discussion}
In summary, we have demonstrated a single photon source in the conventional telecom band from a coupled QD-cavity system. At a resonance, the spontaneous emission rate and the intensity of the single quantum dot were enhanced by the Purcell effect. We therefore obtained a brighter and faster single photon source compared than a single bare QD. This fast (0.2 ns) single photon source has the potential to improve the modulation rate of long-distance quantum telecommunication networks. A simple estimation gives a maximum modulation rate of 5 GHz, which is fast enough for most applications. As for the intensity, the ratio between two PL intensities of the QD at on resonance and out of resonance with the cavity yields a six fold enhancement. This enhancement may improve single photon detection efficiency in the telecom band.

In this study, we collected photons radiated from the cavity, which led to relatively low collection efficiency. The actual photon count rate for the Purcell-enhanced single photon source at the detector is around 950 $\pm$ 190 counts per second (cps).  Although this value is still not sufficient for practical QKD application, we assume that we can largely improve the collection efficiency by placing an output waveguide near the cavity, and also by adding a spot-size converter to the other end of the waveguide, which adiabatically connects a photonic crystal waveguide to a fibre. An alternative approach involves a coupling through a tapered fibre. In fact, one of the most important features of photonic crystals for the present purpose is that a very high extraction efficiency of single photons to an optical fibre can be achieved.

In terms of the coupled $XX$-cavity system, we anticipate that our experimental observation will stimulate theoretical studies of the biexcitonic cavity QED \cite{Oka2008}. On the experimental side, we are aware of the recent observation of the spontaneous two-photon emission of $XX$-cavity coupling in the strong regime, which is interesting as regards the development of heralded single photons and entangled photon pairs \cite{Ota2011}.

\lgchptr{Methods}

\sublgchptr{Device fabrication}
The InAs/InP QD samples were grown by low-pressure metal organic vapor-phase epitaxy using trimethyl-indium, trimethyl-gallium, tertiarybutyl-arsine, and tertiarybutyl phosphide as gas sources. The vicinity of the InP (100) substrates was misoriented by 2 degrees towards (110). The sample structure consisted of 110 nm InP followed by an InAs QD layer plus an ultrathin GaAs monolayer underneath and another 110 nm InP layer placed on a 1500-nm-thick lattice-matched InGaAsP sacrificial layer \cite{Anantathanasarn2005}. The width, height, and area density of the QD are 40 nm, 5 nm, and $\sim$3$\times$10$^{10}$ cm$^{-2}$, respectively. For the PhC fabrication, a periodic hole pattern with a radius of 100 nm arranged in a triangular lattice were defined using electron beam lithography. Cl$_{2}$-based inductively coupled plasma reactive ion etching was used to transfer the pattern to the QD wafer. Finally, we obtained a 220 nm-thick air-bridge photonic crystal with a width-modulated line-defect cavity.

\sublgchptr{Single photon registration with the SSPD}
The photon registration system consisting of a fibre-coupled NbN SSPD with two independent channels was used for time-resolved PL and second-order correlation measurements. We cooled the SSPD system to 1.7 K in a continuous helium vapour flow. The input light was launched into a single-mode optical fibre that was connected to the flange of a cryogenic insert. Then, the SSPD output was connected to the port of the bias current sources and the high-bandwidth, low-noise amplifiers. For our experiments with a 1,550-nm emission wavelength, we optimized the bias current so that it gave a quantum efficiency (QE) of $\sim$4$\%$ and dark count rate of $\le$10 cps.

\lgchptr{Acknowledgment}
We acknowledge stimulating discussions with E. Kuramochi, A. Yokoo, K. Jimyung, K. Nozaki, K. Takeda, and M. Takiguchi. Part of this work was supported by Core Research for Evolutional Science and Technology-Japan Science and Technology Agency (CREST-JST).

\lgchptr{Contributions}
M. D. B. conducted the optical experiments, analysed the data, and wrote the manuscript. H. S. initialised the experiments, built the set-up and designed the structures. S. M. fabricated the sample. H. T. supported the FDTD calculation. P. J. vV. and R. N. fabricated the wafer with InAs/InP QDs. M. N. led the project and partially wrote the paper.

\lgchptr{Competing financial interests}
The authors declare that they have no competing financial interests.

\lgchptr{Corresponding author}
Correspondence to: Masaya Notomi (notomi.masaya@lab.ntt.co.jp).

\newpage

\end{document}